\documentclass[twoside]{article}
\usepackage{fleqn,espcrc2}
\usepackage{epsf}
\usepackage{epsfig}

\usepackage{graphicx}
\usepackage[figuresright]{rotating}

\newcommand{\fnm}[1]{\footnotemark[#1]}

\title{Microscopic theory of the coupling of intrinsic Josephson oscillations 
and phonons}

\author{Ch. Preis\address{Institute of Theoretical Physics, University of
Regensburg, D-93040 Regensburg, Germany}, 

Ch. Helm\fnm{1}\address{Los Alamos National Laboratory, Division T-11, M.S.
B-262, NM 87545, USA},  K. Schmalzl\fnm{1}, Ch. Walter\fnm{1}, J.
Keller\fnm{1}}

\begin{document}

\begin{abstract}
A microscopic theory 
for the coupling of intrinsic
Josephson oscillations and dispersive phonon branches in layered
superconductors is developed. Thereby the effect of phonons on 
the electronic c-axis transport enters through an effective
longitudinal dielectric function.
This coupling provides an explanation of recently observed subgap resonances in
the $I_{\rm dc}$-$V_{\rm dc}$-curve 
of anisotropic cuprate superconductors forming a stack of 
short Josephson junctions. Due to the finite dispersion these resonances can
appear at van-Hove-singularities of both optical and acoustical phonon 
branches,
explaining low-voltage structures in the I-V-characteristic, which are not
understood in phonon models without dispersion.
In long junctions the dispersion of collective electron-phonon 
 modes parallel  to the layers is investigated.
\end{abstract}

\maketitle

%\pacs{74.80Dm, 74.50+r, 74.25Kc, 74.25Jb}

\section{INTRODUCTION}

The main features of the electronic $c$-axis transport  
in the high-$T_c$-superconductors  Tl$_2$Ba$_2$Ca$_2$Cu$_3$O$_{10+\delta}$ 
 and Bi$_2$Sr$_2$CaCu$_2$O$_{8+\delta}$  below 
the critical temperature $T_c$ can be well understood in the model of  
a stack of superconducting  CuO$_2$-layers, 
which form a onedimensional stack of intrinsic Josephson junctions
\cite{kleiner1}.

One of the few phenomena, which are specific to the {\em intrinsic} 
Josephsoneffect and cannot be seen in arrays of conventional superconductors, 
is the interaction of the Josephson oscillations and phonons. This has been 
observed experimentally in the form of resonances in the
$I_{\rm dc}$-$V_{\rm dc}$-characteristics at 
low voltages \cite{schlengal,yurgenss,seidel}. Their main features 
could already 
be understood theoretically in the framework of a simple lattice dynamical 
model of damped local oscillators in the barrier \cite{wir1}.  

Despite the clear evidence of the effect a general multi-band theory, in which 
parameters like the phonon frequencies or the oscillator strength in the 
dielectric function are specified microscopically, is still missing and shall 
be presented in the following. This more general formalism is in principle  
appropriate to include the eigenvalues and eigenvectors of the dynamical matrix
as obtained in {\em any}
 microscopic lattice dynamical calculation into the Josephson
theory. Qualitative effects of the phonon dispersion are discussed in a simple 
rigid ion model, e.g. the van-Hove singularity at 
the upper edge of the acoustical band, which can be detected in the 
$I_{\rm dc}$-$V_{\rm dc}$-curve. Details of the 
derivation and further results can be found elsewhere \cite{wir6}. 

\section{THEORY OF PHONON COUPLING}

%The main steps in the theoretical derivation of the 
%$I_{\rm dc}$-$V_{\rm dc}$-characteristic in short junctions 
%($q_\parallel=0$)  \cite{wir6} shall be briefly 
%summarized in the following. 

The Josephson effect between the layers
$n$ and $n+1$ is described by a RSJ-like equation
$$
j=j_c\sin \gamma_n(t) + j_{qp}(E_n(t)) + \epsilon_0 \dot E^\rho_n(t),
$$
where $j$ is the bias current (density),  
$j_c\sin \gamma_n(t)$ the supercurrent, $j_{qp}(E_n)$  the
quasiparticle current and  $\gamma_n(t)$  the
gauge-invariant phase difference, which is
related to the average electric field $E_n(t)$ in $c$-direction in the barrier:
$\hbar\dot\gamma_n(t)=2edE_n(t)$ ($d$: thickness of the barrier). 
In contrast to this, the displacement
current density $\epsilon_0 \dot E^\rho_n(t)$ does  not  depend on 
$E_n$, but the field 
$E_n^\rho(t)$ set-up by the conduction-electron charge-fluctuations 
$\delta \rho_n$ on the layers, which obey the 
Poisson equation $\delta \rho_n(t)=\epsilon_0(E^\rho(t) - E_{n-1}^\rho(t)$
and the continuity equation 
$j_n (t) - j_{n-1} (t) = - \delta \dot \rho_n (t)$. 
All quantities are constant along the layers ($q_\parallel=0$) 
as in short junctions.

%In short junctions ($q_\parallel=0$) the tunneling current density $j_n$ from
%layer $n$ to $n+1$ creates (two-dimensional) charge density  fluctuations 
%$\delta \rho_n$ on the layers according to the continuity equation 
%\begin{equation}
%j_n (t) - j_{n-1} (t) = - \delta \dot \rho_n (t) \quad . \label{cont}
%\end{equation}
%These charge fluctuations create constant electric fields $E^{\rho}_n$ 
%(in $c$-direction) between the layers, which are given by the Poisson
%equation:                
%\begin{equation}
%\delta \rho_n(t) = \epsilon_0 (E^\rho_n(t) - E^\rho_{n-1}(t) ) \quad . 
%\label{Maxwell1} 
%\end{equation}
%Imposing the bias current density $j$ at the edges,
% the tunneling  current density 
%\begin{equation}
%j = j_c \sin \gamma_n(t) + j_{\rm qp}(E_n(t)) + \epsilon_0 \dot E^{\rho}_n (t)
%\end{equation}
%in each layer $n$ is obtained. 
%%Thereby the usual expression for the Josephson tunneling current has been 
%introduced, where the superconducting phase difference $\gamma_n(t)$ 
%is determined by the average electric field 
%$E_n(t):= \frac{1}{d}\int_{z_n}^{z_{n+1}} E_z(z,t) dz \label{totalE}$ via 
%$\hbar \dot \gamma_n(t) = 2e E_n(t)d$.

The crucial point is to determine the (linear) relation between the field 
\begin{equation}
E^{\rho} ( q_z, \omega)
 = E - E^{\rm ion} = \epsilon_{\rm ph} ( q_z, \omega ) 
   E ( q_z, \omega) 
\end{equation} 
created by the charge fluctuations alone and the  total electric
field $E (q_z, \omega)$ 
due to both electrons and ions. This can be described by the 
(longitudinal) phonon dielectric function 
\begin{equation} 
\epsilon^L_{\rm ph}(q_z, \omega) = \frac{\epsilon_\infty}{1-\chi(q_z, 
\omega)}  \label{ephon} \quad , 
\end{equation}
where the suszeptibility ($\lambda$: phonon branch)
\begin{equation} 
\chi(q_z,\omega) = \sum_\lambda \frac {\vert\Omega(q_z\lambda)\vert^2}
{\omega^2(q_z \lambda) - \omega^2} \label{chi} 
\end{equation} 
and the oscillator strength 
\begin{equation} 
\vert\Omega(q_z\lambda)\vert^2 =
 \sum_{\kappa\kappa'} \tilde Z_\kappa
\frac {e_z(\kappa\vert q_z\lambda) e^*_z(\kappa'\vert q_z\lambda)}
{v_c\epsilon_0 \sqrt{M_\kappa M_{\kappa'}}}\tilde Z^*_{\kappa'}. 
\label{osc} 
\end{equation} 
depend on the eigenvalues $\omega (q_z, \lambda)$ and eigenvectors 
${\vec e} (\kappa\vert q_z\lambda)$ of the dynamical matrix 
$D_{\alpha\beta}({\textstyle{\vec q\atop\kappa \kappa'}})$, which contains 
all the short and long range interaction between the ionic cores in the 
insulating state. 
The eigenvalues $\omega (q_z, \lambda)$ therefore represent 
the {\em bare} phononfrequencies in the absence of the charge fluctuations 
$\delta \rho_n$.
The appearence of special $q_z$-dependent effective charges 
${\tilde Z}_{\kappa}(q_z)$ 
in equ. \ref{osc} reflects the different contribution of
ions on and between the superconducting layers and is one of the distinctions
to the conventional dielectric function. 
The function $\epsilon_{\rm ph}^L (q_z, \omega)$ 
has zeros at the longitudinal eigenfrequencies $\omega( q_z, \lambda)$
of modes with polarization in $c$-direction. 

A similar consideration \cite{wir7} shows the general form 
\begin{equation}
\epsilon_{\rm ph}(\vec q,\omega) = 1 + \frac{\chi(\vec q,\omega)}{1 - f(\vec
q)\chi(\vec q,\omega)} 
\end{equation}
of the phononic dielectric constant, which reduces 
in special cases to the longitudinal 
$\epsilon^L_{\rm ph}(q_z)=\epsilon_{\rm ph}(q_\parallel=0,q_z)$ 
and transversal dielectric constant 
$\epsilon^T_{\rm ph}(q_{\parallel})=\epsilon_{\rm ph}(q_\parallel,q_z=0)=\epsilon^T_{\infty}+\chi(q_\parallel,\omega)$, which is relevant in
 optical experiments.

With the knowledge of the phononic dielectric function $\epsilon_{\rm ph}$ the
$I_{\rm dc}$-$V_{\rm dc}$-curve can be calculated with the ansatz 
\begin{equation}
\gamma_n (t)= 
\left \{
\begin{array}{ll}
\gamma_0 + \omega t + \delta \gamma_n (t) & n \quad {\rm resistive} \\
\gamma_0  + \delta \gamma_n (t) & n \quad {\rm else}
\end{array} , \right.
\end{equation}
by linearizing in the small oscillating parts $\delta \gamma_n (t) \ll 1$.
The final result for the first branch  ($V_{\rm dc}= \hbar \omega/(2e)$, 
$\omega_p^2 := 2edj_c/(\hbar \epsilon_0)$)
\begin{eqnarray}
\label{IV1}
j(V)&=& j_{\rm qp}(V) - 
\frac{j_c}{2} \frac{\omega^2_p}{\omega^2} {\rm Im} \frac
{1}{\tilde \epsilon(\omega)}
%\\ \nonumber
%&=& j_{qp}(V) + \frac{j_c}{2} \frac{\omega^2_p}{\omega^2} \frac
%{\tilde\epsilon_2(\omega)}{\tilde \epsilon_1^2(\omega) + \tilde
%\epsilon_2^2(\omega)}.  
\end{eqnarray}
exhibits a phonon resonance {\em exactly} at the zeros of the modified 
dielectric function
\begin{equation}
\tilde \epsilon(\omega) = 
\Bigl[ \frac{1}{N_z} \sum_{q_z}
\frac{1}{
\epsilon^L_{\rm ph}(q_z,\omega) - \frac{\omega^2_p}{\omega^2} +
\frac{i\sigma}{\epsilon_0 \omega} }  \Bigr]^{-1} \! \! \! \! + 
\frac{\omega^2_p}{\omega^2} ,
\label{epseff}
\end{equation}
i.e. in the unrenormalized phonon bands $\omega ({\vec q}, \lambda)$. 
More specifically peaks develop at the van-Hove Singularities of the 
phonon density of states due to the averaging over $q_z$ in equ. 
\ref{epseff}.
%In special cases these can be suppressed by zeros of the 
%oscillator strength $| \Omega |^2$.

\begin{figure}
%\begin{center}
%\epsfig{bbllx=18,bblly=560,bburx=536,bbury=740,figure=paper3.ps,width=0.5\textwidth,clip=}
%\end{center} 
%\caption{Dispersion and oscillator strengths for the two-atomic
%chain model.}
\epsfig{bbllx=58,bblly=262,bburx=280,bbury=423,figure=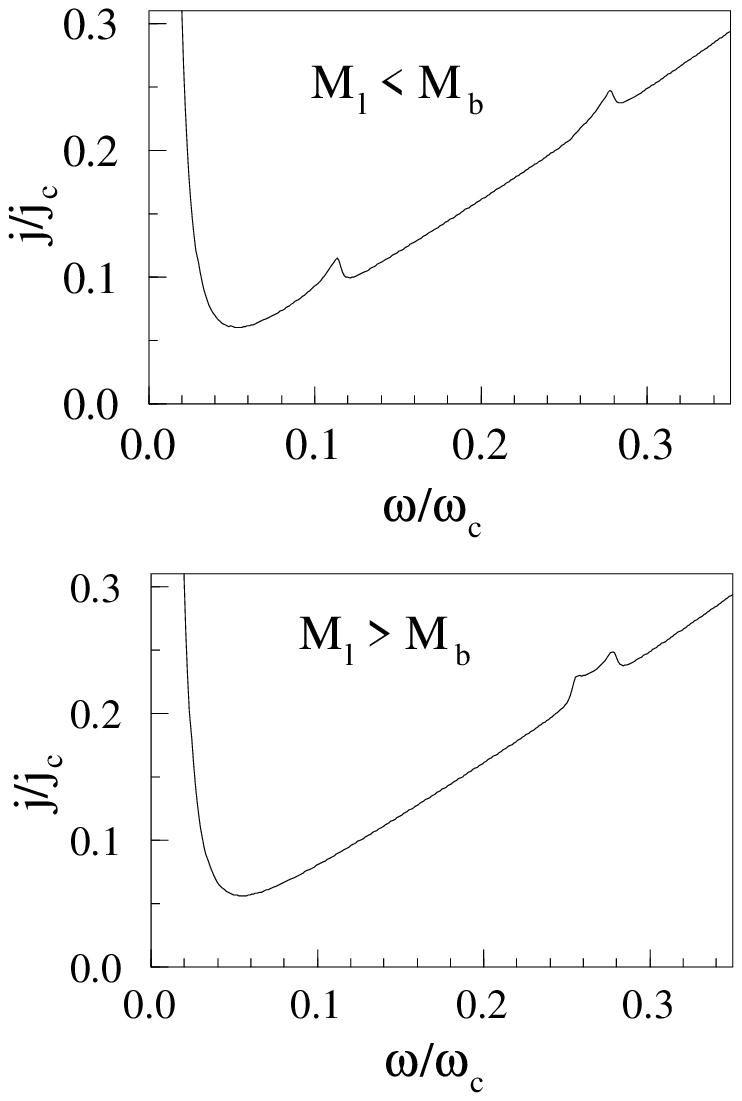,width=0.4\textwidth,clip=}
\caption{$I_{\rm dc}$-$V_{\rm dc}$-curve of the first branch in the 
two-atomic chain model ($M_l>M_b$). \label{fig4}}
\end{figure}  

%\begin{figure}
%\begin{center}
%\epsfig{bbllx=18,bblly=560,bburx=536,bbury=740,figure=paper3.ps,width=0.5\textwidth,clip=}
%\end{center} 
%\caption{Dispersion and oscillator strengths for the two-atomic
%chain model.}
%\epsfig{figure=fig4.eps,width=0.4\textwidth,clip=}
%\caption{ $I$-$V$-curves for one resistive junction with 
%subgap-structures due to acoustical and optical phonons
%for different ratio of the masses on the layer ($M_l$) and in the barrier 
%($M_b$). \label{fig4}}
%\end{figure}  

%Zeros of the real part of this function describe longitudinal collective     
%modes in the system. On the other hand the resonances in the 
%$I_{\rm dc}$-$V_{\rm dc}$-curve
%appear at the {\it bare} longitudinal phonon frequency. The 
%summation over $q_z$ in Eq.\ (\ref{J}) leads to an effective damping of
%the resonances which is proportional to the phonon dispersion. The physical
%origin is the transfer of energy by phonons from the resistive junction to the
%neighboring junctions. 

As a reliable microscopic lattice dynamical theory for the dynamical matrix  
is not yet available and as the intrinsic Josepshon effect has to be 
treated in a phenomenological (one-band) picture until the details of 
the pairing mechanism are understood, the above results will be 
illustrated qualitatively in a simple model of a two-atomic chain of masses
$M_\kappa$ and charges $Z_1=-Z_2$. 

If the mass on the layer ($M_l$) is larger than the one in the barrier 
($M_b$), 
the acoustical van-Hove singularity is suppressed, but the optical band shows
a double peak at the band edges (cf. fig. \ref{fig4}), the distance 
being the bandwidth. 
The (more realistic) case of $M_l<M_b$ exhibits single peaks at the upper 
edges of both acoustical and optical bands and is compared with experiment 
in fig. \ref{fig8}.

If several junctions are in the resistive state, their dynamics is coherent 
due to the coupling with phonons, even if no other (e.g. inductive) 
interaction is present. This is important for the application of short 
junctions as effective high-frequency devices. 
Analytical expressions similar to equ. \ref{IV1} for the second branch 
can in principle distinguish different phase locked solutions in the 
$I_{\rm dc}$-$V_{\rm dc}$-characteristic \cite{wir6}.

\section{COLLECTIVE MODES}

\begin{figure}[htp]
\epsfig{figure=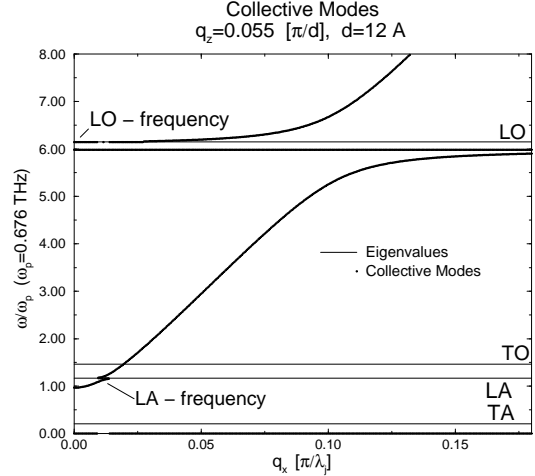,width=0.4\textwidth,angle=-90}
\caption{Dispersion $\omega_{\rm el-ph} (q_x,q_z=0.055)$ of 
collective plasma-phonon modes for $q_x \ll q_z$ (horizontal lines: bare 
phonon frequencies $\omega({\vec q}, \lambda)$, $\lambda_{ab} /d =100$,
 $\lambda_j \approx d \lambda_c/2 \lambda_{ab}$).
 \label{kollfig}   }
\end{figure}

The zeros of the real part of the total dielectric function
\begin{equation}
\epsilon_{\rm tot}({\vec q},\omega) = \epsilon_{\rm ph}({\vec q},\omega) - 
\frac{\omega^2_p ({\vec q}) }{\omega^2} +
\frac{i\sigma}{\epsilon_0 \omega}  ,  \label{epsL}
\end{equation}
define the frequencies $\omega_{\rm el-ph} ({\vec q}, \omega)$ of the 
collective modes of the coupled conduction electrons and phonons,
where the Josephson plasma frequency is given by
($\lambda_{ab}$: coherence length): 
\begin{equation}
\omega_p(\vec q)=\omega_p^2(1+\frac{c_0^2q_{\parallel}^2/\omega_p^2}{1-2\frac{\lambda_{ab}^2}{d^2}(\cos(q_zd)-1)}) .
\end{equation}
These modes
are different from the bare phonon bands $\omega({\vec q},  \lambda)$
and can in principle be observed in neutron scattering. In fig.  
\ref{kollfig} the plasma band $\omega_{\rm pl} ({\vec q})$ for $q_x \ll q_z$ 
mixes with the longitudinal acoustical (LA) and optical (LO) phonon modes 
$\omega ({\vec q},  \lambda)$, which are almost dispersionless on this scale. 
Note especially that a frequency gap of width $\sqrt{2 \omega_{\rm pl} \Omega}$
appears in the spectrum, which is {\em formally} (not physically) 
similar to the polariton dispersion of coupled  phonons and photons
\cite{wir7}. 

\section{EXPERIMENTAL RESULTS}

Recently the explanation of the subgap resonances in 
Refs.\cite{schlengal,yurgenss,seidel,wir1} with  the 
phonon coupling mechanism presented here could be well confirmed by 
Raman measurements on the same samples \cite{yurgenss,yurgenspriv}
and infrared reflectivity experiments with grazing 
incidence \cite{tsvetkov,tajima} (see table \ref{phonon_tab}). 
%Small deviations of the order of $\sim 5-10 \%$ can be attributed to the 
%fact that in optical experiments and in the intrinsic Josephson effect
%different averages over $\vec q$ of the dielectric functions are relevant.
Note that in our theory also Raman-active or silent modes may
couple to intrinsic Josephson oscillations for $q_z\ne0$ and that 
in contrast to optical experiments (at the $\Gamma$-point) 
the whole Brioullin zone contributes in an average way 
(cf. equ. \ref{epseff}). 
This provides  a reason for slight discrepancies of these 
experimental data in table \ref{phonon_tab} 
%With the help shell-model calculations
%\cite{kulkarnitl1} some of the more pronounced 
%structures can be connected with certain elongation patterns of the ions in 
%the unit cell. 
%For example, the peak in the $I$-$V$-curve at 4.64 THz in 
%Tl$_2$Ba$_2$Ca$_2$Cu$_3$O$_{10}$ seems to be due to a (Cu,Ba)-mode. 
For more experimental and theoretical references see \cite{wir6,tsvetkov}.

\begin{table}[htp]
\caption{  Comparison 
of the frequencies $f_{\rm sg}=h~V_{\rm dc}/2e$
(in THz) of the most pronounced subgap resonances 
and of infrared- ($f_{LO}$) and Raman active ($f_{TO}$)
modes in Bi$_2$Sr$_2$CaCu$_2$O$_8$ and Tl$_2$Ba$_2$Ca$_2$Cu$_n$O$_{2n+4}$.   
\label{phonon_tab}}
    \begin{tabular}{ccccccc}
    \hline
    \multicolumn{6}{|c|}{  Phonons in
      Bi$_2$Sr$_2$CaCu$_2$O$_8$ } \\ 
    \hline   
    $f_{\rm sg}$ & 2.97 & 3.89 & 5.17 & 5.60 &  \cite{wir1} 
    
    \\ \hline
    $f_{\rm LO}$ & 2.85 &      & 5.07 &      &  \cite{tsvetkov} 
    \\ \hline
    $f_{\rm LO}$  & 2.86 &      & 5.16 &      &  \cite{tajima}  \\ \hline
    $f_{\rm TO}$ &      & 3.80 &      &     
    &  \cite{yurgenss,yurgenspriv}  \\ 
      \hline
     \multicolumn{6}{|c|}{Phonons in
       Tl$_2$Ba$_2$Ca$_2$Cu$_n$O$_{2n+4}$
      } \\ \hline 
    $f_{\rm sg}$ & 3.63 & 4.64 & & &  \cite{wir1} $n=3$ 
     \\ \hline
     $f_{\rm LO}$ &      & 4.50 & & &   \cite{tsvetkov} $n=2$
     \end{tabular}
%     \footnotetext[1]{Tl$_2$Ba$_2$Ca$_2$Cu$_3$O$_{10}$}
%     \footnotetext[2]{Tl$_{2}$Ba$_{2}$Ca$_{2}$Cu$_{2}$O$_{8}$}
\end{table}

The qualitative features of the subgap resonances can already be understood in 
the local oscillator model of \cite{wir1}: The 
position of the resonance  is completely independent  on temperature, magnetic 
field or the geometry of the probe, while the intensity of the structure 
varies $\sim j_c^2$ with the critical current density $j_c(T,B)$. 
Also the behaviour in external pressure is consistent with the phonon 
interpretation \cite{yurgenspriv}.

\begin{figure}
\epsfig{bbllx=18,bblly=50,bburx=552,bbury=460,figure=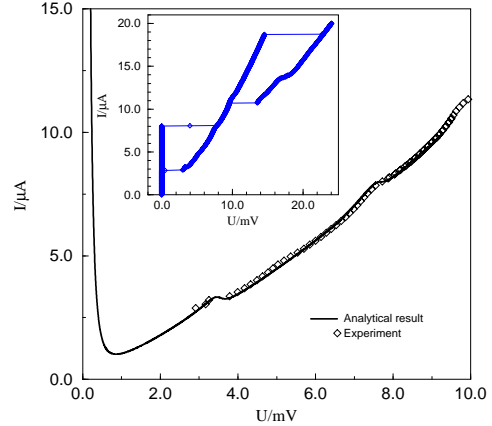,width=0.49\textwidth,clip=}
\caption{
Fit of the experimental $I_{\rm dc}$-$V_{\rm dc}$-curve in 
Tl$_2$Ba$_2$Ca$_2$Cu$_3$O$_{10}$ \protect\cite{seidel} near
the band edge of the acoustical branch (at $1.5$ THz) and an optical 
branch within the two-atomic chain model (inset: first two resistive branches).
\label{fig8}}
%\vspace{-0.05\textheight}
\end{figure}

Beyond this, in the more general model developed above resonances at 
van-Hove singularities, e.g.  at the upper band edge of the 
acoustical phonon band, can be described.  This might be an explanation for
 a peak seen in \cite{seidel} at $3.2$ mV ($\hat{=} 1.54$ THz) 
in the $I_{\rm dc}$-$V_{\rm dc}$-characteristic of 
Tl$_2$Ba$_2$Ca$_2$Cu$_3$O$_{10}$, as
the same frequency is predicted in lattice dynamical calculations
\cite{kulkarnitl1} for the upper edge of the acoustical band  
and because there are no optical phonon bands at that low frequency
(cf. Fig. \ref{fig8}).
Also the double peaks shown in fig. \ref{fig4} might have
been seen in the satellite structures at 5.17 THz (10.7 mV) and 5.6 
(11.6 mV) in the $I_{\rm dc}$-$V_{\rm dc}$-curve 
of Bi$_2$Sr$_2$CaCu$_2$O$_{8+\delta}$ \cite{wir1}, which is consistent with 
the theoretical width $\sim 0.3$ THz of this phonon band 
\cite{kulkarnitl1}.

\section{CONCLUSIONS} 

In this contribution the  microscopic theory for the coupling between
Josephson oscillations and longitudinal phonons in intrinsic Josephson 
systems like the highly anisotropic cuprate superconductors
has been developed. 
 The analytical result equ. (\ref{IV1}) for 
the $I_{\rm dc}$-$V_{\rm dc}$-curve for one resistive junction and
the detailed form of the longitudinal dielectric function equ. (\ref{ephon})
describing the 
coupling has been  obtained. 
In paritcular, not only optical but also acoustical phonons at the 
edge of the Brillouin zone couple to Josephson oscillations
explaining a structure observed in \cite{seidel} in the 
$I_{\rm dc}$-$V_{\rm dc}$-curve occurring at low voltage.
The discussion of the dispersion of collective phonon-electron modes is an 
important first step to understand the effect of phonons in long junctions
\cite{wir7}.

%Josephson oscillations and phonons. For this purpose a simplified lattice
%dynamical model has been used with one acoustical and one optical branch. 
%It is found that in the   limit of large values of the McCumber parameter
%the numerical results follow closely the analytical
%solutions with the following exceptions: 1. Using a gradual change of the
%bias-current, regions of the $I$-$V$-curve with negative differential
%conductivity are skipped as is observed in current-biased experiments.
%2. In the case of several resistive junctions, where several analytical
%solutions are obtained, the numerical result
%follows one of the analytical solutions. The stability of the different
%analytical solutions is currently investigated. It seems to be that the
%solution which gives a minimum for the interaction energy between
%polarisation and the electric field generated by the Josephson oscillations
%at a given frequency is most stable.
%The phonons thus serve to synchronise the Josephson oscillations in 
%different  
%resistive layers, which is important for the application of such systems as
%high-frequency devices.   

The authors thank A. Yurgens, A. Tsvetkov, A. Mayer, D. Strauch,
R. Kleiner, P. M{\"u}ller, L. Bulaevskii and A. Bishop for fruitful 
discussions and DFG, FORSUPRA and DOE under contract W-7405-ENG-36 (C.H.) 
for financial support.


\begin{thebibliography}{99}
\bibitem{kleiner1} R. Kleiner et al., 
       Phys. Rev. Lett. {\bf 68}, 2394 (1992); R. Kleiner et al., 
      Phys. Rev. B {\bf 49} (1994), 1327.
\bibitem{schlengal} K. Schlenga, et al., 
      Phys. Rev. Lett. {\bf 76}, 4943 (1996).
\bibitem{yurgenss}
     A. Yurgens et al., Proceedings of SPIE, Vol {\bf 2697}, 433 (1996).
\bibitem{seidel} P. Seidel et al., 
     Physica C {\bf 293},  49 (1997).
\bibitem{wir1} Ch.~Helm et al., 
%Ch.~Preis, F.~Forsthofer, J.~Keller,
%     K.~Schlenga, R.~Kleiner,  P.~M\"uller, 
      Phys. Rev. Lett. {\bf 79}, 737 (1997);
%\bibitem{wir2}
%Ch.~Preis, F.~Forsthofer, J.~Keller,
%     K.~Schlenga, R.~Kleiner,  P.~M\"uller, 
   Ch.~Helm et al.,    Physica C {\bf 293}, 60 (1997);
%\bibitem{wir3} K. Schlenga et al., 
% R. Kleiner, G. Hechtfischer,
%     M. M\"o{\ss}le, S. Schmitt, P. M\"uller, Ch. Helm, Ch. Preis,
%     F. Forsthofer, J. Keller, H. L. Johnson, M. Veith, E. Steinbei{\ss},
  K. Schlenga et al.,     Phys. Rev. B {\bf 57}, 14518 (1998). 
%\bibitem{Zett} T. Zetterer, M. Franz, J. Sch\"utzmann, W. Ose, H.H. Otto, K.F.
%     Renk, Phys. Rev, B 41, 9499 (1990); V.M. Burlakov, S.V. Shulga, J.
% Keller,
%     K.F. Renk, Physica C {\bf 203}, 68 (1992) and references therein. 
%\bibitem{Buck} R.G. Buckley, M.P. Staines, D.M.  Pooke, T. Stoto, N.E. Flower,
%     Physica C {\bf 248}, 247 (1995) and references therein. 
\bibitem{wir6} Ch. Helm et al.,  
% Ch. Preis, Ch. Walter, J. Keller, 
submitted to Phys. Rev. B,
 cond-mat/9909318.  
\bibitem{wir7} Ch. Preis, K. Schmalzl, Ch. Helm, J. Keller, in preparation.
%\bibitem{kinder} P. Berberich, R. Buemann, H. Kinder, 
%       Phys. Rev. Lett.  {\bf 49},   1500 (1982).
%\bibitem{Pon} Ya.G. Ponomarev, E.B. Tsokur, M.V. Sudakova, S.N. Tchesnokov,
%      M.A. Lorenz, M.A. Hein, G. M\"uller, H. Piel, B.A. Animov, preprint. 
%\bibitem{Koy96} T. Koyama, M. Tachiki,
%      Phys. Rev. B {\bf 54}, 16183 (1996).  
%\bibitem{wir4} Ch. Preis, Ch. Helm, J. Keller, A. Sergeev, R. Kleiner,       
%      SPIE Conference 
%     Proceedings ``Superconducting Superlattices II: Native and Artificial'',
%     Vol {\bf 3480}, 236 (1998).
%\bibitem{wir5} Ch. Preis et al, to be published.
%\bibitem{Maksimov} E.G. Maksimov, P.I. Arseev, N.S. Maslova, preprint cond-mat
%     9812422.
%\bibitem{Zeyher} R. Zeyher, G. Zwicknagl, Z. Phys. B {\bf 78},
%     175 (1990).
\bibitem{yurgenspriv} A. Yurgens, private communication. 
\bibitem{tsvetkov} A. A.
     Tsvetkov et. al.,
%D. Duli\'{c}, D. van der Marel, A.    Damascelli, G. A.
%     Kaljushnaia, J. I. Gorina, N. N. Senturina,   B. Willemsen, N. N.
%     Kolesnikov, Z. F. Ren, J. H. Wang,  A. A. Menovsky,    T. T. M. Palstra,
     preprint. 
\bibitem{tajima} S. Tajima et al., 
%G. D. Gu, S. Miyamoto, A. Odagawa, N. Koshizuka, 
       Phys. Rev. B {\bf 48}, 16164  (1993). 
%\bibitem{jia} C.-S. Jia, P.-Y. Lin, Y. Xiao, X.-W. Jiang, X.-Y. Gou, 
%     S. Huo, H. Li, Q.-B. Yang, 
%        Physica C {\bf 268}, 41 (1996).
\bibitem{kulkarnitl1} A. D. Kulkarni et al., 
%J. Prade, F. W. de Wette, W. Kress, U. Schr{\"o}der, 
%Phys. Rev. B {\bf 40}, 2642 (1989);  
%A. D. Kulkarni et al., 
%F. W. de Wette, J. Prade, U. Schr{\"o}der, W. Kress, 
Phys. Rev. B {\bf 43}, 5451 (1991).
%J. Prade et al., 
%A. D. Kulkarni, F. W. de Wette, U. Schr{\"o}der, W. Kress, 
%Phys. Rev. B {\bf 39}, 2771 (1989).

\end{thebibliography}
\end{document}